\documentclass[showpacs]{revtex4}

\usepackage[dvips]{graphicx}
\input{epsf}
\usepackage{amsmath}
\usepackage{amssymb}
\usepackage[mathscr]{eucal}
\usepackage{bm}
\usepackage{subfigure}

\newcommand{\beq}{\begin{equation}}
\newcommand{\eeq}{\end{equation}}
\newcommand{\beqa}{\begin{eqnarray}}
\newcommand{\eeqa}{\end{eqnarray}}
\newcommand{\beqan}{\begin{eqnarray*}}
\newcommand{\eeqan}{\end{eqnarray*}}

\newcommand{\tr}[1]{{\rm tr} \left( #1 \right) }
\newcommand{\ket}[1]{| #1 \rangle}
\newcommand{\bra}[1]{\langle #1 |}

\newcommand{\llangle}{\langle\!\langle} 
\newcommand{\rrangle}{\rangle\!\rangle}

\newcommand{\proof}{\noindent {\bf Proof. }}
\newcommand{\qed}{\hfill $\Box$ \vskip 2ex}

\newtheorem{proposition}{Proposition}

\newtheorem{corollary}{Corollary}

\begin{document}

\title{Tensor of coherences parameterization of  multiqubit density operators for entanglement characterization}
\author{Claudio Altafini}
\affiliation{SISSA-ISAS  \\
International School for Advanced Studies \\
via Beirut 2-4, 34014 Trieste, Italy }
\email{altafini@sissa.it}


\begin{abstract}
For multiqubit densities, the tensor of coherences (or Stokes tensor) is a real parameterization obtained by the juxtaposition of the affine Bloch vectors of each qubit.
While it maintains the tensorial structure of the underlying space, it highlights the pattern of correlations, both classical and quantum, between the subsystems and, due to the affine parameterization, it contains in its components all reduced densities of all orders.
The main purpose of our use of this formalism is to deal with entanglement.
For example, the detection of bipartite entanglement is straightforward, as it is the synthesis of densities having positive partial transposes between desired qubits.
In addition, finding explicit mixtures for families of separable states becomes a feasible issue for few qubit symmetric densities (we compute it for Werner states) and, more important, it provides some insight on the possible origin of entanglement for such densities.

\end{abstract}

\maketitle

\section{Introduction}

The main purpose of this work is to discuss a parameterization of a multiqubit (pure or mixed) density operator and to show its usefulness in dealing with classical and quantum correlations.
The principle behind the parameterization is the same of the so-called vector of coherences of widespread use in modeling $N$-level density operators, \cite{Alicki1,Cla-contr-open1,Byrd2,Kimura1,Tilma1}.
It consists in choosing a complete orthogonal set of Hermitian operators and in considering the corresponding real vector of expectation values in place of the density matrix.
Here we use the same idea, but respecting the tensorial structure of the density, hence working with a (real) {\em tensor of coherences}.
Each qubit is parameterized as an affine Bloch vector, and the tensor is just the juxtaposition of affine Bloch vectors.
Maintaining the tensorial structure has several advantages, for example it makes the pattern of the ``total'' correlation between subsystems totally straightforward to see.
For ``total'' correlation we mean both the classical and the quantum ones.
As a matter of fact, we will see that the correlation between subsystems is encoded in the terms of the tensor.
Crucial to the understanding of this point is the role of the affine component and of how it enters into the compounding of the different qubits. 
In fact, as we use homogeneous coordinates to deal with the affine term, the key simplification is that tracing over one of the qubits simply corresponds, up to a scale factor, to choosing the ``0'' (i.e., affine) component for the corresponding index.
Hence, because of the affine parameterization, reduced densities are naturally represented by means of the tensor of coherences parameterization and the tensor itself consists of the entire hierarchy of correlations.
Also the scale factor has a natural interpretation: it corrects the trace norm of the completely random state when passing from a density to a reduced density.
Further advantages are in the simplicity of the geometric picture for multipartite systems and in the possibility of using multilinear algebra ideas in a more straightforward manner.

Obviously the parameterization we consider is not really new; it was treated in detail in \cite{Fano2}, used extensively for example by Mahler and co-workers \cite{Mahler1,Schlienz1} (where the basis elements we use are referred to as cluster operators) or in \cite{Jaeger1,James1} (where it is referred to as the Stokes tensor) and more or less implicitly in many other papers, cf. \cite{Byrd2,Jakobczyk1} for related material.
For example in the NMR literature \cite{Ernst1} it goes under the name of product of operators basis. What is new is its use in understanding multiparty entanglement, see \cite{Eckert1,Horodecki6,Lewenstein1,Terhal1} for an overview of research in this field.

The simplest (to detect) type of entanglement is bipartite entanglement, for which there exist a necessary and sufficient condition, the so-called positive partial transpose (PPT) criterion of \cite{Peres1,Horodecki1}. 
In the tensor of coherences parameterization, the PPT criterion has a very simple formulation and, more important, it becomes completely trivial to construct densities satisfying PPT between all pairs of subsystems. 
Hence one can focus on the class of entangled PPT densities, which are characterized by the more subtle bound entanglement \cite{Horodecki2}. 
On the other hand, also the construction of systems having certain patterns of bipartite entanglement not satisfying the PPT criterion (NPT entanglement) is rather simple.

If the total correlation, classical plus quantum, is directly depicted in the tensor of coherences, the distinction between the two types of correlation remains however an elusive issue, although as we will see in the examples, the parameterization allows to suggest what is happening in an entangled state. 
Consider a one-parameter family of densities ending in the maximally mixed state. Close to such extreme the state is certainly separable \cite{Terhal1}. 
Using the tensor of coherences it is not too difficult to construct an explicit (one parameter) convex mixture of product states for it.
In all the examples of entangled families we have tested, the convex combination found is such that it induces cancellations between the corresponding mixtures of reduced densities. 
If close enough to the complete mixing these cancellations are harmful, far from it it may happen that the one parameter density is well-defined while some of the reduced densities (which, again, being canceled do not explicitly appear unless one wants to construct the mixture explicitly) are not anymore well-defined densities in the sense that their trace norm is too big (some of its eigenvalues become greater than 1). 
Hence one source of entanglement is that not well-defined components give rise to a well-defined compound system.
The complication is obviously that due to the nonuniqueness of the mixture representing a given density, it is an hard problem to exclude that any other convex combination will suffer from the same ``unfeasibility'' problem.
For low rank systems, the tensorial notation helps in finding such convex combinations. For example we could easily compute a mixture for the Werner states valid in the whole separability interval.

The parameterization into vector (and tensor) of coherences is natural only for qubits. For a $k$-level system, in fact, the vector of coherences of the density operator is not free to evolve on the corresponding (affine) ball in $ \mathbb{R}^{k^2-1} $, see \cite{Byrd2,Kimura1} for hints on this point.
Of course qubits are by far the most popular systems in quantum information processing.

One may argue that the dimension of the state tensor grows as $ 2^{2n} $ with the number $n$ of qubits and hence that expanding densities explicitly into a complete basis becomes rapidly cumbersome.
The exponential growth of the number of degrees of freedom available concerns however all densities, regardless of the representation used.  
While this fact is immediately evident using our notations, it may go unnoticed using some standard parameterization. 
Of course in a problem like detecting entanglement all degrees of freedom of the state may come into play, therefore we find it convenient to have them all explicitly expressed.

In next Section the tensor of coherences is introduced and correlations, both classical and quantum, are discussed in its terms.
For sake of notation simplicity, we consider in some detail the geometry of the 2-qubit case.
The extension to $n$-qubit densities is straightforward. 
In Section~\ref{sec:examples} several examples are treated.
We construct explicit mixtures for Werner states and for a tripartite family ending into the bound entangled state of \cite{Bennett1}.
An example of how to construct (and analyze) NPT tripartite entanglement is also proposed.

\section{Tensor of coherences parameterization for $\rho$}
Given $ n$ qubits living on the Hilbert space $ ({\cal H}^2)^{\otimes n}$ of dimension $ 2^n $, the corresponding density operator is a $ 2^n \times 2^n $ positive semidefinite Hermitian matrix $ \rho$ such that $ \tr{\rho}=1 $ and it has $ 2^{2n} -1 $ degrees of freedom.
We construct for $ \rho $ a basis borrowed from the literature on NMR spectroscopy where it is normally referred to as the product of operator basis \cite{Ernst1}. Similar bases are discussed for example in \cite{Fano2,Jaeger1,James1,Schlienz1}.
In terms of this basis, studying densities is equivalent to studying tensors of directly observable real parameters.

A word on the notation: we use the symbol ``$ \rho $'' for density matrices and ``$ \varrho^{j} $'' (possibly with a multiindex) for the components of the tensor of coherences.
The superindex is always a tensor (multi)index; for powers of $ \varrho^j $ we use an extra round bracket.
For the tensor, we also use the summation convention over repeated indexes, always in the range $ \{ 0, \ldots, 3 \} $.

\subsection{One qubit}

The rescaled Pauli matrices $ \lambda_j = { \scriptstyle \frac{1}{\sqrt{2}} } \sigma_j$, $ j=1, \, 2 , \, 3$,
\[
\lambda_1 = \frac{1}{\sqrt{2}}  \begin{bmatrix} 0 & 1 \\ 1 & 0 \end{bmatrix} \quad
\lambda_2 = \frac{1}{\sqrt{2}}  \begin{bmatrix} 0 & -i \\ i & 0 \end{bmatrix} \quad
\lambda_3 = \frac{1}{\sqrt{2}}  \begin{bmatrix} 1 & 0 \\ 0 & -1 \end{bmatrix}
\]
plus the rescaled identity operator $ \lambda_0 = \frac{1}{\sqrt{2}} \openone_{2\times 2} $ form a complete orthonormal basis (in the sense that $ \tr{ \lambda_j \lambda_k} = \delta_{jk} $) for $ 2\times 2 $ Hermitian matrices. Fixing the trace means fixing the component along $ \lambda_0 $.
Hence $ \rho $ can be expressed as the affine 3-vector
\[
\rho =\varrho^0 \lambda_0 + \varrho^1 \lambda_1 + \varrho^2 \lambda_2 + \varrho^3 \lambda_3 = \varrho^j \lambda_j 
\]
where $ \varrho^j = \tr{\rho \lambda_j } $, $ j=1, \, 2, \, 3 $ and the component along $ \lambda_0 $ is $ \varrho^0 = \tr{\rho \lambda_0 }= \tr{\rho}/\sqrt{2} =1/\sqrt{2} $. 
Since $ \varrho^0 $ is a constant, it is normally neglected and only the Bloch vector $ \vec{\varrho} = ( \varrho^1 \; \varrho^2 \; \varrho^3 )^T $ is considered. 
However, here it is convenient to keep the constant part and to represent the affine vector in terms of a set of homogeneous coordinates, i.e., by means of the 4-vector $ \bar{\varrho}= ( \varrho^0 \; \varrho^1 \; \varrho^2 \; \varrho^3 )^T $.
From $ \tr{\lambda_j \lambda_k } = \delta_{jk} $, $  j, \, k =0 , \, 1, \, 2, \, 3 $, $ \tr{\rho_1 \rho_2} $ induces an inner product on the parameter space (${\mathbb{R}^4 } $) given by $ \tr{\rho_1 \rho_2 } =  \llangle \bar{\varrho}_1 , \, \bar{\varrho}_2 \rrangle =  \varrho_1^0\varrho_2^0  +  \llangle \vec{\varrho}_1 , \, \vec{\varrho}_2 \rrangle =  \frac{1}{2}  +  \llangle \vec{\varrho}_1 , \, \vec{\varrho}_2 \rrangle  $. The norm of $ \bar{\varrho} $ is then given by $ \| \bar{\varrho} \|= \sqrt{ \llangle \bar{\varrho} , \, \bar{\varrho} \rrangle} = \sqrt{\tr{\rho^2 } }   = \sqrt{ \frac{1}{2}  +  \| \vec{\varrho} \|^ 2 } =  \sqrt{\frac{1}{2}  + r^2 } $.
Purity corresponds to $ \tr{\rho^2} =1 $ i.e., $ \| \vec{\varrho} \| ^2 =\frac{1}{2} $ or $ \vec{\varrho} $ belonging to the sphere of radius $r = \frac{1}{\sqrt{2}} $, call it $ \mathbb{S}^2 _{1/\sqrt{2} } $, while the 4-vector $ ( \varrho^0 \;  \varrho^1 \; \varrho^2 \; \varrho^3 )^T $ belongs to the affine sphere $ \left( \varrho^0, \; \mathbb{S}^2_{1/\sqrt{2} } \right)=  \left( \frac{1}{\sqrt{2}},  \; \mathbb{S}^2_{1/\sqrt{2} } \right) \subset \mathbb{S}^3_{1}  $.
Complete mixing, given by $ \rho =\frac{1}{2} \openone_{2\times 2}= \frac{ \sqrt{2}}{2} \lambda_0 $, has norm $ \tr{\rho^2  } = (\varrho^0 )^2 = \frac{1}{2} $ and corresponds to $ \vec{\varrho} = \vec{0} $ i.e., to a ``sphere'' of 0 radius.
All degrees of mixing are in between the two extremes just presented and in general the Bloch vector $ \vec{\varrho} \in \mathbb{S}^2 _{r} $ for $ 0 \leqslant r \leqslant \frac{1}{\sqrt{2}} $.
Hence we have $ - \frac{1}{2} \leqslant  \llangle \vec{\varrho}_1 , \, \vec{\varrho}_2 \rrangle  \leqslant \frac{1}{2} $ and $ 0 \leqslant  \llangle \bar{\varrho}_1 , \, \bar{\varrho}_2 \rrangle  \leqslant 1 $, $ \forall \;  \vec{\varrho}_1 , \, \vec{\varrho}_2 \in \mathbb{S}^2_{r} $.

\subsection{Two qubits}
\label{sec:two-qubits}

Call $ \Lambda_{jk} = \lambda_j \otimes \lambda_k $, $  j, \,k \in \{ 0, \, 1 ,\, 2,\, 3\}  $.
Up to a normalization constant, the $\Lambda_{jk} $ form the so-called {\em product operator basis}, see \cite{Ernst1}, and are subdivided into 
\beqan 
\text{ $0$ qubit operators}  & & \Lambda_{00} \\
\text{ $1$ qubit operators}  & & \Lambda_{01}, \,\Lambda_{02}, \,\Lambda_{03}, \,\Lambda_{10}, \,\Lambda_{20}, \,\Lambda_{30}  \\
\text{ $2$ qubit operators}  & & \Lambda_{11}, \, \Lambda_{12}, \,\Lambda_{13}, \,\Lambda_{21}, \,\Lambda_{22}, \,\Lambda_{23}, \,\Lambda_{31} , \, \Lambda_{32}, \,\Lambda_{33} 
\eeqan
Similarly to the 1-qubit case, the set of $  \Lambda_{jk} $ $ j, \, k \in \{ 0, \, 1, \, 2,\, 3 \} $ forms an orthogonal basis for all $ 4 \times 4 $ Hermitian matrices (as $\rho$ is now). It is still normalized i.e., such that $ \tr{\Lambda_{jk}\Lambda_{lm} }= \tr{\lambda_j \lambda_l \otimes \lambda_k \lambda_m }= \tr{\lambda_j \lambda_l}  \tr{\lambda_k \lambda_m }=  \delta_{jl}\delta_{km} $ for all $ j, \, k, \, l, \, m \in \{ 0, \, 1, \, 2, \, 3 \} $.
Except for $ \Lambda_{00}$, every $ \Lambda_{jk} $ has 2 eigenvalues $ \pm \frac{1}{2} $, each with multiplicity 2. 
An equivalent description of $ \rho$ is given by the 2-tensor $ \varrho^{jk} $, $ j, \, k = 0, \ldots 3,$ where $ \varrho^{jk}= \tr{\rho \Lambda_{jk} }$ i.e.,
\beq
\rho = \varrho^{jk} \Lambda_{jk} = \varrho^{jk} \lambda_j \otimes \lambda_k 
\label{eq:two-rho}
\eeq
The {\em tensor of coherences} $ \varrho^{jk} $ can still be seen as the 16-vector $ [ \varrho^{00} \, \varrho^{01} \ldots \varrho^{03} \, \varrho^{10} \ldots \varrho^{33} ] ^T $, which is still a homogeneous representation of an affine 15-vector since $ \varrho^{00} $ is a constant.
In fact, as $ \Lambda_{jk} $ is traceless for $ \{ jk\} \neq \{ 00\} $ and $ \Lambda_{00} =  \frac{1}{2} \openone_{2\times 2} \otimes \openone_{2\times 2} $ has trace 2, $\tr{\rho} =1 $ implies that $ \varrho^{00} = \tr{\rho \Lambda_{00} } = \tr{\rho}/2 =  1/2 $.
Again the $ \varrho^{jk} $ parameterization lives on $ \mathbb{R}^{16} $ endowed with the  Euclidean inner product one gets from the following:
\beq
\tr{\rho^2}  = \tr{ \left( \varrho^{jk} \Lambda_{jk}\right)^2 } 
= \sum_{j, k=0}^3 \left( \varrho^{jk} \right)^2 
= {\rm const} \leqslant 1
\label{eq:Casim1}
\eeq
Following the terminology of \cite{Byrd2}, the norm $ \tr{\rho^2} $ is a quadratic Casimir invariant of $ \rho$.
Following instead for example \cite{Rossignoli1}, $ \tr{\rho^2 } $ is a Tsallis entropy corresponding to the choice of parameter $ q=2 $ in $ S_q(\rho) = \left( \tr{\rho^q} - 1\right)/(1-q)$.

Unlike the single qubit case, the subset of $ \mathbb{R}^{16} $ in which the parameters $ \varrho^{jk} $ are such that $ \varrho^{jk} \Lambda_{jk} $ is a well-defined density operator is not at all clear a priori and a hierarchy of nested subsets exists: 
\begin{quote}
uncorrelated $ \subset $ separable $ \subset $ entangled $ \subset $ ``nondensity'' $ \subset $ $ \mathbb{R}^{16}$.
\end{quote}

If the density operator $ \rho$ is {\em uncorrelated} (i.e. $ \rho $ is a product state: $ \rho = \rho_A \otimes \rho_B $), then also $ \varrho^{jk} $ can be intended as the tensor product (which for scalar quantities becomes ordinary multiplication)
\beq
\varrho^{jk} = \varrho^j_A \otimes \varrho^k_B = \varrho^j_A \varrho^k_B
\label{eq:tens-prod-rho}
\eeq
with $ \varrho^j_A $ describing the state of the first spin and $ \varrho^k_B $ the state of the second.
By ``fully stretching'' $ \varrho^j_A \otimes \varrho^k_B $, one obtains still the $16$-dimensional vector.
Eq. \eqref{eq:tens-prod-rho} is not true for correlated states: for example $\bar{\varrho} = [\frac{1}{2} \, 0 \, \ldots  \, 0 \,  \varrho^{33} ] $ has no expression of the form \eqref{eq:tens-prod-rho}.
Notice that since $ \varrho^0_A = \varrho^0_B $ are nonzero constants, if $ \varrho^{jk} =0 $ for all pairs $ \{ j k \} $ such that $ j\neq 0 $ and $ k\neq 0$, then $ \rho $ is uncorrelated.
From the same argument, it follows that, even for correlated densities, it is always possible to write $ \varrho^{j0} $ and $ \varrho^{0k} $ in the form: 
\beq
  \varrho^{j0} = \varrho^j_A \otimes \varrho^0_B =\frac{1}{\sqrt{2}} \varrho^j_A \quad \text{and} \quad   
\varrho^{0k} = \varrho^0_A \otimes \varrho^k_B =\frac{1}{\sqrt{2}} \varrho^k_B . 
\label{eq:0-separat}
\eeq
The corresponding $ \varrho^j_A =\sqrt{2} \varrho^{j0} $ and $ \varrho^k_B=
\sqrt{2} \varrho^{0k} $ are univocally determined.
Therefore $ \varrho^{00} = \varrho^0_A \otimes \varrho^0_B = \frac{1}{2}$, regardless of the uncorrelation of $ \rho$.

The main difference with respect to other papers like \cite{Fano2,Schlienz1} is that we include the $ 0$-qubit and $1$-qubit terms in the dyadic tensor structure.
As a consequence, in the basis \eqref{eq:two-rho}, the reduced density operator is very natural to obtain as the partial trace operation consists simply in selecting the component of index ``0'' in the qubit to be traced over.

\begin{proposition}
\label{prop:reducedd}
Given $ \rho = \varrho^{jk} \Lambda_{jk}$, the reduced density operator is given by 
\begin{subequations}
\beqa
\label{eq:partial-tr}
\rho_A & = &  {\rm tr}_B \left( \rho \right)=\varrho^j_A \lambda_j = \sqrt{2} \varrho^{j 0 } \lambda_j  
\\
\rho_B & =& {\rm tr}_A \left( \rho \right) =  \varrho^k_B \lambda_j= \sqrt{2} \varrho^{0 k} \lambda_k  .
\eeqa
\end{subequations}
\end{proposition}

\proof
Since $ \tr{\lambda_k } = \delta_k $, 
\[
\rho_A = {\rm tr}_B \left( \rho \right) = \varrho^j_A \lambda_j = \varrho^{jk} {\rm tr}_B \left( \Lambda_{jk} \right) = \varrho^{jk} \lambda_j \otimes {\rm tr} \left( \lambda_{k} \right) = \sqrt{2} \varrho^{j 0 } \lambda_j 
\]
or $ \varrho^j_A  =  \sqrt{2} \varrho^{j0} $ 
from \eqref{eq:0-separat}.
Similarly for $ \rho_B $.
\qed

In general (also for noncorrelated or nonseparated densities) the numbers $ \tr{\rho_A^2} $ and $ \tr{\rho_B^2 } $ in \eqref{eq:Casim-tens} are partial quadratic Casimir invariants i.e, the quadratic Casimir invariants of the two reduced densities.
The scale factor $ \sqrt{2} $ on all components of the reduced density plays a double ``normalization'' role: it takes care of the trace and it modifies the quadratic Casimir invariant of the completely mixed state (which changes with the number of qubits).

For an uncorrelated $\rho$, from the tensor product structure, $ \varrho^{jk} = \varrho^j_A \otimes \varrho^k_B $ is living on the tensor product of two affine spheres in $ \mathbb{R}^4 $:
\beq
\varrho^j_A \otimes \varrho^k_B \in \left(  \frac{1}{\sqrt{2}}, \; \mathbb{S}^2_{r_A } \right) \otimes  \left(   \frac{1}{\sqrt{2}}, \;  \mathbb{S}^2_{r_B } \right) , \qquad 0  \leqslant r_A, \, r_B \leqslant \frac{1}{\sqrt{2}} 
\label{eq:manifold}
\eeq
and therefore eq. \eqref{eq:Casim1} becomes:
\beq
\begin{split}
\tr{\rho^2}
& =   \sum_{j, k=0}^3 \left(\varrho^j_A \otimes \varrho^k_B  \right)^2 
 =\sum_{j, k=0}^3 (\varrho^j_A )^2\otimes ( \varrho^k_B  )^2 \\
& =  \left( \frac{1}{2} + \| \vec{\varrho}_A \|^2 \right) \otimes  \left( \frac{1}{2} + \| \vec{\varrho}_B \|^2 \right) 
=  \left( \frac{1}{2} + r_A ^2 \right)  \left( \frac{1}{2} + r_B ^2 \right) = \tr{ \rho_A ^2 } \tr{\rho_B^2 }  .
\end{split}
\label{eq:Casim-tens} 
\eeq
Under local orthogonal transformations, $ r_A ={\rm const} $ and $ r_B = {\rm const } $. Under nonlocal rotations instead \eqref{eq:Casim-tens} is not anymore valid.
Likewise, the factorization is not valid for correlated $ \rho$.

Equation \eqref{eq:manifold} is useful to have a geometric picture of the manifold on which $ \varrho^{jk} $ lives.
Expanding into its 16 components, we have 
\begin{itemize}
\item 
 $ \varrho^{00} =\frac{1}{2} $ is the affine part;
\item $ \left( \varrho^{10} \, \varrho^{20} \,\varrho^{30} \right) \in  \mathbb{S}^2_{r_A  /\sqrt{2}} $;
\item $ \left( \varrho^{01} \, \varrho^{02} \,\varrho^{03} \right) \in  \mathbb{S}^2_{r_B  /\sqrt{2}} $;
\item the remaining 9-dimensional vector $  \left( \varrho^{11} \, \varrho^{12} \,\varrho^{13} \,  \varrho^{21} \, \varrho^{22} \,\varrho^{23} \, \varrho^{31} \, \varrho^{32} \,\varrho^{33}  \right) \in  \mathbb{S}^2_{r_A } \otimes \mathbb{S}^2_{r_B } $.
\end{itemize}

With the parameterization chosen, testing correlation becomes an almost tautological issue. For example, a necessary condition for uncorrelation of $\rho$ is that $ \varrho^{j0} \neq 0 $ and $ \varrho^{0k} \neq 0 $ $ \, \forall \, \varrho^{jk} \neq 0 $, $ j, \, k = 1, \, 2, \, 3$.
In fact, assume that $ \rho $ is uncorrelated and $ \varrho^{jk} \neq 0 $ for some pair $ \{ jk\} $, $ j, \, k = 1, \, 2, \, 3$. 
If for example $ \varrho^{j0} =0 $ then from $ \varrho^{j0} = \varrho^j_A \otimes \varrho^0_B $ we must have that $ \varrho^j_A = 0 $ and hence $  \varrho^{jk} = \varrho^j_A \otimes \varrho^k_B = 0$.
Another necessary condition for uncorrelation of $ \rho $ is that $ \tr{ \rho^2_A } \tr{\rho^2_B} = \tr{\rho^2} $.
That these conditions are not sufficient for uncorrelation is easily shown by examples, see Section~\ref{sec:ex-an-example}.
Following \cite{Fano2}, it is convenient to describe the correlation between A and B by means of a {\em correlation tensor} $ {\cal C} = c^{jk} \Lambda_{jk} $ (homogeneous i.e, such that $ c^{0k} = c^{j0} =0 $) and rewrite $ \rho $ as 
\beq
\rho = \rho_A \otimes \rho_B + {\cal C} 
 = \left( \varrho^j_A \lambda_j \right) \otimes \left( \varrho^k_B \lambda_k \right) + c^{jk} \Lambda_{jk} 
\label{eq-2qubit-corr}
\eeq
where $ c^{jk} = \varrho^{jk} - \varrho^j_A \varrho^k_B $, $ j,k=1, \ldots ,3$.
Hence the necessary and sufficient condition for uncorrelation of $\rho $ is that $ {\cal C} =0 $.
On the other extreme, when the reduced densities $ \rho_A $ and $ \rho_B $ are completely mixed the tensor of coherences contains only correlations: $ \varrho^{jk} = c^{jk} $, $ j,k=1, \ldots , 3$.
From \eqref{eq-2qubit-corr} we can compute $ \| \bar{\varrho} \|^2 =  \| \bar{\varrho}_A \|^2 \| \bar{\varrho}_B \|^2 + \sum_{j,k=1}^3 \left( c^{jk}\right)^2  $, obtaining the classical inequality 
\beq
 \tr{\rho^2} \leq \tr{\rho_A^2} \tr{\rho_B^2} 
\label{eq:class-tr2-ineq}
\eeq

A classically correlated or {\em separable} state is written as a linear convex combination \footnote{In \eqref{eq:conv-comb-2-qubits}, the summation over $ p$ is in the range $ \{ 1, \ldots ,s \} $ and is obviously intended with respect to both components of the tensor product $ \rho = \sum_{p=1}^s  w^p \left( \rho_{A, p} \otimes \rho_{B,p} \right) $.}
\beq
 \rho = w^p \rho_{A, p} \otimes \rho_{B,p}, \qquad  w^p \geqslant 0 , \qquad \sum_{p=1}^s w^p =1 .
\label{eq:conv-comb-2-qubits}
\eeq
If for pure states separability is equivalent to uncorrelation, for mixed states the simultaneous presence of both classical and quantum correlations complicates considerably the picture.

The convex combination reduces under partial trace operation and the statement of Proposition \ref{prop:reducedd} implies

\begin{proposition}
\label{prop:reduc-dens}
Given $ \rho = w^p \rho_{A, p} \otimes \rho_{B,p} $, $ w^p \geqslant 0 $, $ \sum_{p=1}^s w^p =1 $, the reduced density operators are given by 
\begin{subequations}
\beqa
\label{eq:partial-tr-sep}
\rho_A & = &  {\rm tr}_B \left( \rho \right) = w^p \rho_{A, p} = \sqrt{2} w^p \varrho^{j0}_{A,p} \lambda_j  
\label{eq:partial-tr-sep-1}
\\
\rho_B & =&  {\rm tr}_A \left( \rho \right)  = w^p \rho_{B, p}  = \sqrt{2} w^p \varrho^{0k}_{B,p} \lambda_k  .
\label{eq:partial-tr-sep-2}
\eeqa
\end{subequations}
\end{proposition}

\proof
The result is well-known and the proof is reported here as an exercise in computing with the tensor of coherences.
Considering $ \rho $ 
\[
\begin{split}
\rho & = \left( w^1 \varrho^0_{A,1} \varrho^0_{B,1} + \ldots + w^s \varrho^0_{A,s} \varrho^0_{B,s} \right) \Lambda_{00} +  \left( w^1 \varrho^0_{A,1} \varrho^1_{B,1} + \ldots + w^s \varrho^0_{A,s} \varrho^1_{B,s} \right) \Lambda_{01} + \ldots + \\ 
& +\left( w^1 \varrho^3_{A,1} \varrho^3_{B,1} + \ldots + w^s \varrho^3_{A,s} \varrho^3_{B,s} \right) \Lambda_{33}
\end{split}
\]
and tracing over $B$
\beqa
\rho_A & = & {\rm tr}_B \left( \rho \right) = \sqrt{2} \left( \left(  w^1 \varrho^0_{A,1} \varrho^0_{B,1} + \ldots + w^s \varrho^0_{A,s} \varrho^0_{B,s} \right) \lambda_{0} +  \right.\nonumber \\
&&  \left.+ \left(  w^1 \varrho^1_{A,1} \varrho^0_{B,1} + \ldots + w^s \varrho^1_{A,s} \varrho^0_{B,s} \right) \lambda_{1} + \ldots  + \left(  w^1 \varrho^3_{A,1} \varrho^0_{B,1} + \ldots + w^s \varrho^3_{A,s} \varrho^0_{B,s} \right) \lambda_{3} \right)\nonumber \\
& = &  \sqrt{2} \left( w^1 \left( \varrho^0_{A,1} \lambda_0 +\ldots + \varrho^3_{A,1} \lambda_3 \right) \varrho^0_{B,1}  + \ldots +  w^s  \left( \varrho^0_{A,s} \lambda_0 +\ldots + \varrho^3_{A,s} \lambda_3 \right)\varrho^0_{B,s} \right) \nonumber\\
& = &  w^1 \rho_{A,1} + \ldots +  w^s \rho_{A,s} 
\label{eq:reduc-rho-proof}
\eeqa
since $ \varrho^0_{B,p}=\frac{1}{\sqrt{2}} $, $ \forall \, p=1, \ldots, s$.

\qed

It is possible to expand $ \rho = w^p \rho_{A, p} \otimes \rho_{B,p} $ emphasizing its affine structure as follows. 
\beqa
\rho & = &  \frac{1}{2} \Lambda_{00} + \frac{1}{\sqrt{2}} w^p \varrho_{A,p}^j \Lambda_{j0}  + \ldots + \frac{1}{\sqrt{2}} w^p \varrho_{A,p}^3 \Lambda_{30} +  \frac{1}{\sqrt{2}} w^p \varrho_{B,p}^1 \Lambda_{01} + \ldots +  \frac{1}{\sqrt{2}} w^p \varrho_{B,p}^3 \Lambda_{03} \nonumber \\
& & + w^p \varrho_{A,p}^1 \varrho_{B,p}^1  \Lambda_{11} + \ldots +  w^p \varrho_{A,p}^3 \varrho_{B,p}^3  \Lambda_{33} \nonumber \\
& = &  \frac{1}{2} \Lambda_{00} +  \rho_A \otimes  \frac{1}{\sqrt{2}} \lambda_0 +  \frac{1}{\sqrt{2}} \lambda_0 \otimes \rho_B + w^p \varrho_{A,p}^1 \varrho_{B,p}^1  \Lambda_{11} + \ldots +  w^p \varrho_{A,p}^3 \varrho_{B,p}^3  \Lambda_{33} .
\label{eq:rho-sep-from-red}
\eeqa
The expression \eqref{eq:rho-sep-from-red} allows to easily provide a geometric picture of the state space of $ \rho $, extending the one for uncorrelated densities. If $ r_A = w^p r_{A,p} $ and $  r_B = w^p r_{B,p} $, then from \eqref{eq:rho-sep-from-red}:
\begin{itemize}
\item $  \frac{1}{2} $ is the affine component
\item $ \left( \varrho^{10} \, \varrho^{20} \,\varrho^{30} \right)=  \frac{1}{\sqrt{2}} \vec{\varrho}_A  \in  \mathbb{S}^2_{r_A  /\sqrt{2}} = w^p \mathbb{S}^2_{r_{A,p}  /\sqrt{2}}$
\item $ \left( \varrho^{01} \, \varrho^{02} \,\varrho^{03} \right)= \frac{1}{\sqrt{2}} \vec{\varrho}_B  \in  \mathbb{S}^2_{r_B  /\sqrt{2}}= w^p \mathbb{S}^2_{r_{B,p}  /\sqrt{2}} $
\item $\left( \varrho^{11} \,\ldots \,\varrho^{33} \right)= \left(  w^p \varrho_{A,p}^1 \varrho_{B,p}^1 \, \ldots \,  w^p \varrho_{A,p}^3 \varrho_{B,p}^3  \right) \in  w^p   \mathbb{S}^2_{r_{A,p} } \otimes \mathbb{S}^2_{r_{B,p} } $.
\end{itemize}
Notice how in the last item the ``multiplicative'' part of $ \rho $ lives on a convex sum of tensor products of spheres.
Varying $ r_{A,p}$, $ r_{B,p}$ and $ w^p $, all convex combinations of tensor products of spheres of all possible radii can be obtained.
Since any point in the Bloch ball is a good vector of coherences for a qubit and since a convex combination of compact convex sets is compact and convex, the ``multiplicative'' part of a separable $ \rho $, $ \varrho^{11}, \ldots, \varrho^{33} $, lives on a compact, convex, infinitely generated set \footnote{A convex set is said infinitely generated if the set of its extreme points, i.e., the points being expressed as convex combinations only in a trivial manner, is infinite.}.

For $ \rho $ separable, $ \rho = w^p \rho_{A, p} \otimes \rho_{B,p} $, we have
\beq
\begin{split}
\tr{\rho^2} & = \sum_{p,q=1}^s w^p w^q \llangle \bar{\varrho}_{A,p}\, , \,  \bar{\varrho}_{A,q} \rrangle  \llangle \bar{\varrho}_{B,p} \, , \,  \bar{\varrho}_{B,q} \rrangle \\
& = (w^p)^2 \| \bar{\varrho}_{A,p}\| ^2 \| \bar{\varrho}_{B,p}\| ^2  + \sum_{\scriptsize \begin{array}{c} p,q=1,\\ p\neq q \end{array}}^s w^p w^q \llangle \bar{\varrho}_{A,p}\, , \,  \bar{\varrho}_{A,q} \rrangle  \llangle \bar{\varrho}_{B,p} \, , \,  \bar{\varrho}_{B,q} \rrangle.
\end{split}
\label{eq:tr-rho^2-sep}
\eeq
$  \llangle \bar{\varrho}_{A,p}\, , \,  \bar{\varrho}_{A,q} \rrangle $ is an inner product between vectors of possibly different lengths: $  \llangle \bar{\varrho}_{A,p}\, , \,  \bar{\varrho}_{A,q} \rrangle = \frac{1}{2} + \llangle \vec{\varrho}_{A,p}\, , \,  \vec{\varrho}_{A,q} \rrangle $ with $ \vec{\varrho}_{A,p} \in \mathbb{S}^2_{r_{A,p} } $, $   \vec{\varrho}_{A,q}\in \mathbb{S}^2_{r_{A,q} } $.
Since $ 0 \leqslant r_{A, j} \leqslant \frac{1}{\sqrt{2}} $, $ j=p, \,q $, we have by the Cauchy-Schwarz inequality $ - \frac{1}{2} \leqslant - r_{A, p} r_{A, q} \leqslant \llangle \vec{\varrho}_{A,p}\, , \,  \vec{\varrho}_{A,q} \rrangle \leqslant r_{A, p} r_{A, q} \leqslant \frac{1}{2} $ and hence $  0 \leqslant \llangle \bar{\varrho}_{A,p}\, , \,  \bar{\varrho}_{A,q} \rrangle \leqslant 1 $.

For $ \rho $ separable, beside \eqref{eq:class-tr2-ineq} we also recover the necessary condition of \cite{Wu1} (itself a consequence of the partial disorder criterion, see \cite{Nielsen2}) on the traces of the reduced densities 
\begin{proposition}
\label{prop:nec-tr-2}
Given $ \rho = w^p \rho_{A, p} \otimes \rho_{B,p} $, $ w^p \geqslant 0 $, $ \sum_{p=1}^s w^p =1 $, we have that 
\begin{subequations}
\label{eq:tr-sq-red}
\beqa
\tr{\rho_A^2} & \geqslant & \tr{\rho^2} 
\label{eq:tr-sq-red1}
\\
\tr{\rho_B^2} & \geqslant & \tr{\rho^2} .
\label{eq:tr-sq-red2}
\eeqa
\end{subequations}
\end{proposition}

\proof

From Proposition~\ref{prop:reduc-dens},
\beq
\begin{split}
\tr{\rho_A^2} & =\sum_{p,q=1}^{s}  w^p w^q \llangle \bar{\varrho}_{A,p} , \,  \bar{\varrho}_{A,q} \rrangle \\
& =  (w^p)^2 \| \bar{\varrho}_{A,p}\| ^2  + \sum_{\scriptsize \begin{array}{c} p,q=1,\\ p\neq q \end{array}}^s w^p w^q \llangle \bar{\varrho}_{A,p} , \,  \bar{\varrho}_{A,q} \rrangle .
\end{split}
\label{eq:tr-rhoA^2-sep}
\eeq
Equation \eqref{eq:tr-rho^2-sep} and \eqref{eq:tr-rhoA^2-sep} have the same number of terms, but each term of \eqref{eq:tr-rhoA^2-sep} is greater or equal than the corresponding one in \eqref{eq:tr-rho^2-sep}, since $  \| \bar{\varrho}_{B,p}\| ^2 \leqslant 1 $ and $ \llangle \bar{\varrho}_{B,p}\, , \,  \bar{\varrho}_{B,q} \rrangle \leqslant 1 $.
\qed

Eq. \eqref{eq:tr-sq-red} could be rewritten in terms of the tensor of coherences as 
\beq
\| \bar{\varrho}_A \|^2 \geqslant \| \bar{\varrho} \|^2, \qquad 
\| \bar{\varrho}_B \|^2 \geqslant \| \bar{\varrho} \|^2.
\label{eq:trace-ineq-norm}
\eeq
From \eqref{eq:trace-ineq-norm}, the geometric interpretation of \eqref{eq:tr-sq-red} is straightforward: $ \| \bar{\varrho}_A \|^2 \geqslant \| \bar{\varrho} \|^2 $ implies $ \frac{1}{2} + r_A^2 \geqslant  \frac{1}{4} + r^2 $ i.e., the radius of the sphere of the compound system is bounded above by those of the reduced densities: $ 0 \leqslant r^2 \leqslant {\rm min} \{ r_A^2, \, r_B^2 \} +  \frac{1}{4} $.
The coefficient $ \frac{1}{4} $ takes care of the displacements of the centers of the $ \rho$ and $ \rho_A $ (or $ \rho_B $) spheres due to the different affine terms.
Such property is invariant to convex combinations.
Notice that this is a necessary but not sufficient condition for separability (indeed it is not even sufficient for uncorrelatedness) of $ \rho$. 
It becomes a necessary and sufficient condition if $ \rho $ is pure, see \cite{Wu1}.

Also the Partial Transpose operation of \cite{Peres1} $\rho^{T_1}  =  \left( T \otimes \openone_{2} \right) \left( \rho \right) $ and $ \rho^{T_2}  =  \left( \openone_{2} \otimes  T \right) \left( \rho \right) $ (where $T$ denotes the single qubit transposition) becomes a very straightforward operation in the chosen basis.

\begin{proposition}
\label{prop:PPT2}
The partial transpositions of $ \rho$ become the sign changes on the corresponding index ``2'':
\begin{subequations}
\label{eq:PPTvectch}
\beqa
\rho^{T_1} & = & \varrho^{0k} \Lambda_{0k} +  \varrho^{1k} \Lambda_{1k} -  \varrho^{2k} \Lambda_{2k} +  \varrho^{3k} \Lambda_{3k} 
\label{eq:partT1vectch}\\
\rho^{T_2} & = & \varrho^{j0} \Lambda_{j0} +  \varrho^{j1} \Lambda_{j1} -  \varrho^{j2} \Lambda_{j2} +  \varrho^{j3} \Lambda_{j3} .
\label{eq:partT2vectch}
\eeqa
\end{subequations}
\end{proposition}
The proof is by direct computation.
As is well-known \cite{Horodecki1}, the PPT criterion provides a necessary and sufficient condition for separability in the two qubit case: $ \rho $ is separable if and only if $ \rho^{T_1} $ (and $ \rho^{T_2} $) is a density. We say in this case that $ \rho $ is PPT.
Proposition~\ref{prop:PPT2} provides the following constructive separability condition.

\begin{corollary}
\label{cor:PPT-index-2}
A sufficient condition for PPT is that all terms of indexes '2' in the tensor $ \varrho^{jk} $ are 0.
\end{corollary}

For the tensor $ \varrho^{jk} $, the partial transposition is a linear, norm preserving operation: $ \tr{ \rho^2}  =  \tr{ \left(\rho^{T_1} \right)^2}=  \tr{ \left(\rho^{T_2} \right)^2} $. 
Hence entanglement violating PPT does not modify the quadratic Casimir invariants of the density and the necessary conditions \eqref{eq:tr-sq-red} are insensible to it.

\subsection{$n$ qubits}

The notation for a density operator composed of $ n$ qubits (labeled $ A_1, \ldots, A_n $) is completely analogous to the two-qubit case: 
\[
\rho= \rho_{A_1 \ldots A_n} = \varrho^{j_1 \ldots j_n} \Lambda_{j_1 \ldots j_n} = \varrho^{j_1 \ldots j_n} \lambda_{j_1} \otimes \ldots \otimes \lambda_{j_n}, \qquad   j_1, \ldots,  j_n = 0, \, 1, \, 2, \, 3
\]
If $ \Lambda_{j_1 \ldots j_n} $ has $ k \leqslant n $ nonnull indexes, then it is an operator on $ k$ qubits. 
The object $ \varrho^{j_1 \ldots j_n} $ can still be seen as an $n$-tensor or as a $ 2^{2n}-1 $ affine vector having constant component $  \varrho^{0 \ldots 0}= \varrho^0_{A_1} \otimes \ldots \otimes \varrho^0_{A_n }  = \left( \frac{1}{\sqrt{2}} \right)^n $ (check that indeed $ \tr{\Lambda_{0 \ldots 0 } } = \tr{\lambda_0}^n =  \left(\sqrt{2} \right)^n $, hence $ \tr{\varrho^{0 \ldots 0 } \Lambda_{0 \ldots 0 } } =1 $).
Reduced density operators are obtained analogously to the bipartite case by collapsing each index being traced over to $0$ and rescaling by $ \sqrt{2}$ each time.
For example, the reduced density operator of the $k$-th qubit is obtained by tracing over the other $n-1$ qubits:
\[
\rho_{A_k} = {\rm tr}_{A_1 \ldots A_{k-1} A_{k+1} \ldots A_n} ( \rho_{A_1 \ldots A_n} ) = \varrho^{j_k} _{A_k} \lambda_{j_k} = (\sqrt{2})^{n-1} \varrho^{0 \ldots 0 j_k 0 \ldots 0 } \lambda_{j_k}, \qquad j_k=0, \ldots, 3 
\]
Similarly to the bipartite case, we have that $ \rho $ is said uncorrelated if 
\[
\rho = \rho_{A_1} \otimes \ldots \otimes \rho_{A_n } = \varrho^{j_1}_{A_1} \otimes \ldots \otimes \varrho^{j_n}_{A_n } \lambda_{j_1}  \otimes \ldots \otimes \lambda_{j_n} = \varrho^{j_1}_{A_1}\varrho^{j_2}_{A_2} \ldots \varrho^{j_n}_{A_n } \Lambda_{j_1 j_2 \ldots j_n} ,
\]
and for $ \rho $ uncorrelated we have 
\[
\tr{\rho^2} = \tr{\rho_{A_1} } \tr{\rho_{A_2} } \ldots \tr{\rho_{A_n} } = 
\left( \frac{1}{2} + \| \vec{\varrho}_{A_1} \|^2 \right) \left( \frac{1}{2} + \| \vec{\varrho}_{A_2} \|^2 \right) \ldots \left( \frac{1}{2} + \| \vec{\varrho}_{A_n} \|^2 \right) 
\]
A necessary and sufficient condition for uncorrelation of $\rho $ is that for all $n$-tuples of indexes $ \{ j_1 \ldots j_n \} $, $ j_1 , \ldots, j_n = 0, \, 1, \, 2 ,\, 3$
\beq
\varrho^{j_1 \ldots j_n} = \varrho^{j_1}_{A_1} \otimes \ldots \otimes  \varrho^{j_n}_{A_n} = \varrho^{j_1}_{A_1} \ldots \varrho^{j_n }_{A_n} \qquad \forall \;  j_1, \ldots,  j_n = 0, \, 1, \, 2, \, 3
\label{eq:n-qubit-nec-suff}
\eeq
Obviously the test \eqref{eq:n-qubit-nec-suff} can be restricted to $n$-tuples $\{ j_1 \ldots j_n \} $ such that at least two indexes are nonzero.

The density $ \rho $ is said separable if 
\beq
\rho = w^p \rho_{A_1, p} \otimes \ldots \otimes \rho_{A_n,p} 
\label{eq:separ-n}
\eeq
where $ \sum_{p=0}^s  w^p = 1 $, $ w^p \geqslant 0 $.
Unlike the 2-qubit case, no neat necessary and sufficient condition for separability is known, although for few qubits complete classifications in various entanglement classes have been proposed, for example in \cite{Acin1}.
Qualitatively, one distinguishes between entanglement due to violation of the PPT test for some of the indexes in $  \{ j_1 \ldots j_n \} $ and a more subtle type of entanglement, called bound entanglement which cannot be detected by means of PPT.
The first type of entanglement is bipartite and will be referred to as NPT entanglement.

\begin{proposition}
\label{prop:NPTentangl-n}
$ \rho $ is NPT entangled if for some $ k \in  \{ 1 \ldots n \} $ the partial transpose 
\beq
\begin{split}
\rho^{T_k} & = \varrho^{j_1 \ldots j_{k-1} 0 j_{k+1} \ldots j_n } \Lambda_{j_1 \ldots j_{k-1} 0 j_{k+1} \ldots j_n }
+  \varrho^{j_1 \ldots j_{k-1} 1 j_{k+1} \ldots j_n } \Lambda_{j_1 \ldots j_{k-1} 1 j_{k+1} \ldots j_n } \\
&
- \varrho^{j_1 \ldots j_{k-1} 2 j_{k+1} \ldots j_n } \Lambda_{j_1 \ldots j_{k-1} 2 j_{k+1} \ldots j_n }
+ \varrho^{j_1 \ldots j_{k-1} 3 j_{k+1} \ldots j_n } \Lambda_{j_1 \ldots j_{k-1} 3 j_{k+1} \ldots j_n }
\end{split}
\label{eq:PT-n-qubits}
\eeq
has negative eigenvalues.
\end{proposition}
Also Corollary \ref{cor:PPT-index-2} still holds.
\begin{corollary}
\label{cor:PPT-index-n}
A sufficient condition for PPT is that all terms of indexes '2' of the tensor $ \varrho^{j_1 \ldots j_n } $ are 0.
\end{corollary}
Similarly, if $ \pi(\cdot ) $ is a permutation of $ 1, \ldots, n $, the necessary condition of Proposition~\ref{prop:nec-tr-2} generalizes to the sequence of inequalities \cite{Wu1}
\beq
\tr{ \rho_{A_{\pi(1)} }^2  } \geqslant \tr{ \rho_{A_{\pi(1)} A_{\pi(2)}}^2  } \geqslant  \ldots \geqslant\tr{ \rho^2  } \qquad \forall \; \text{ permutations }\pi(\cdot )
\label{eq:tr-norm-ineq-n}
\eeq

\section{Examples}
\label{sec:examples}

\subsection{Partial quadratic Casimir invariants and classical correlations: an example}
\label{sec:ex-an-example}
The aim of this example is to show that given a 2-qubit density $ \rho$, the condition $ \tr{\rho^2} = \tr{\rho_A^2} \tr{ \rho_B^2} $ is not a sufficient condition for uncorrelation.
Consider the bipartite density: $ \rho = \varrho^{00} \Lambda_{00} +  \varrho^{01} \Lambda_{01} +  \varrho^{03} \Lambda_{03} +  \varrho^{10} \Lambda_{10} +  \varrho^{30} \Lambda_{30} +  \varrho^{13} \Lambda_{13} + \varrho^{31} \Lambda_{31} $.
Once $ \varrho^{01}$, $ \varrho^{03}$, $ \varrho^{10}$ and $ \varrho^{30} $ are fixed, then $  \varrho^{1}_A$, $ \varrho^{3}_A$, $ \varrho^{1}_B$ and $ \varrho^{3}_B $ are determined.
Choose $ \varrho^{13} $ and $  \varrho^{31} $ such that 
\beqan 
 \varrho^{1}_A\varrho^{3}_B & \neq &  \varrho^{13} \\
 \varrho^{3}_A\varrho^{1}_B & \neq & \varrho^{31} 
\eeqan
but such that 
\[
(\varrho^{1}_A)^2 (\varrho^{3}_B)^2 +(\varrho^{3}_A)^2 (\varrho^{1}_B )^2 = (\varrho^{13})^2 + (\varrho^{31} )^2
\]
Then $ \tr{\rho^2_A} \tr{\rho^2_B } = \tr{\rho^2} $ but $ {\cal C} \neq 0 $ implying that this state is correlated.
Since $ \varrho^{2k} = \varrho^{j2} =0$, from Corollary~\ref{cor:PPT-index-2} the correlation is classic.

\subsection{Bell state}
\label{sec:ex-Bell}
The Bell state $ \ket{\psi} = \frac{1}{\sqrt{2}} \left( \ket{0} + \ket{1} \right) $ has density operator given by
\[
\rho_{\rm Bell}= \frac{1}{2} \begin{bmatrix} 1 & 0 & 0 & 1 \\
0 & 0 & 0 & 0 \\
0 &  0 & 0 & 0 \\
1 & 0 & 0 & 1
\end{bmatrix}
\]
In the tensor of coherences, it corresponds to 
\[
\rho_{\rm Bell}= \varrho^{jk} \Lambda_{jk} =  \frac{1}{2} \Lambda_{00} - \frac{1}{2}\Lambda_{11} - \frac{1}{2}\Lambda_{22} + \frac{1}{2}\Lambda_{33} 
\]
Since $ \varrho^{j0} = \varrho^{0k} = 0 $ $ \forall \, \{ j k \} \neq \{ 00\}$, this state has both reduced densities $ \rho_A $ and $\rho_B $ that are completely mixed $ \tr{\rho^2_A }= \tr{\rho^2_B}=\frac{1}{2} $. Hence the Bloch vectors $\vec{\varrho}_A $ and $\vec{\varrho}_B $ live on ``spheres'' of 0 radius and in \eqref{eq-2qubit-corr} $ \varrho^{jk} = c^{jk} $, $ j,k =1, \ldots, 3 $.
$  \tr{\rho^2_{\rm Bell}  } = 1 $ implies that the entangled state instead is pure.

\subsection{Werner states}
\label{sec:ex-Werner}
Consider the one-parameter family of states
\beq
\rho_{\rm Wer}(x) = \begin{bmatrix} \frac{1-x}{4} & 0 & 0 & 0 \\
0 & \frac{1+x}{4} & - \frac{x}{2} & 0 \\
0 &  - \frac{x}{2} &  \frac{1+x}{4} & 0 \\
0 & 0 & 0 & \frac{1-x}{4} 
\end{bmatrix}
\label{eq:Wer-dens}
\eeq
which is known to be maximally entangled when $ x=1 $ and uncorrelated when $ x=0$.
Using the computations of Appendix~\ref{app:two-spin-comp}, we get the following nonnull components: $ \varrho^{00} = \frac{1}{2}$, $\varrho^{11}  =  - \frac{x}{2} $, $\varrho^{22}  =  - \frac{x}{2} $ and $ \varrho^{33}  =  - \frac{x}{2} $.
Hence
\beq
 \rho_{\rm Wer} (x) = \frac{1}{2} \Lambda_{00} - \frac{x}{2}\Lambda_{11} - \frac{x}{2}\Lambda_{22} - \frac{x}{2}\Lambda_{33} 
\label{eq:werner}
\eeq
For $ x=0 $, $\tr{\rho^2_{\rm Wer}(0) } = \frac{1}{4} $ and, for $ x=1 $, $ \tr{\rho^2_{\rm Wer}(1) } =1 $.
From \eqref{eq:werner}, the reduced density operators $ \rho_A $ and $ \rho_B $ are both completely mixed states $ \forall \, x \in [ 0, \, 1 ] $, since $\varrho^{j0}=\varrho^{0k} = 0 $ $ \forall \, \{ j k \} \neq \{ 00\}$, so, once again, the homogeneous part of $ \rho_{\rm Wer} $ contains only ``pure'' correlations: $ {\cal C} = \rho_{\rm Wer} - \frac{1}{2} \Lambda_{00} $.

For this highly symmetric density, the explicit computation of a linear combination of tensor products which is separable in the whole domain $ 0 \leqslant x \leqslant \frac{1}{3} $ is rather easy with the insight given by the tensor of coherences.
For example, we can construct 2-qubit terms $ \varrho^{jj} $ such that the corresponding reduced terms $ \varrho^{j0} $ and $ \varrho^{0j} $ are identically $0$ in the following way. 
Choose the 12 one-qubit densities:
\beqan
 \rho_{A,1} = \varrho^0 \lambda_0 + \varrho^1_{A,1} \lambda_1 & \qquad
 \rho_{A,2} = \varrho^0 \lambda_0 + \varrho^2_{A,2} \lambda_2 & \qquad
 \rho_{A,3} = \varrho^0 \lambda_0 + \varrho^3_{A,3} \lambda_3 \\
 \rho_{A,4} = \varrho^0 \lambda_0 - \varrho^1_{A,1} \lambda_1 & \qquad
 \rho_{A,5} = \varrho^0 \lambda_0 - \varrho^2_{A,2} \lambda_2 & \qquad
 \rho_{A,6} = \varrho^0 \lambda_0 - \varrho^3_{A,3} \lambda_3 
\eeqan
\beqan
 \rho_{B,1} = \varrho^0 \lambda_0 + \varrho^1_{B,1} \lambda_1 & \qquad
 \rho_{B,2} = \varrho^0 \lambda_0 + \varrho^2_{B,2} \lambda_2 & \qquad
 \rho_{B,3} = \varrho^0 \lambda_0 + \varrho^3_{B,3} \lambda_3 \\
 \rho_{B,4} = \varrho^0 \lambda_0 - \varrho^1_{B,1} \lambda_1 & \qquad
 \rho_{B,5} = \varrho^0 \lambda_0 - \varrho^3_{B,2} \lambda_2 & \qquad
 \rho_{B,6} = \varrho^0 \lambda_0 - \varrho^3_{B,3} \lambda_3 
\eeqan
If we use equal weights $ w^1 = \ldots w^6 =\frac{1}{6} $, the resulting density $ \rho = w^p \rho_{A,p} \otimes \rho_{B,p} $ is
\beqa
\rho &= &  \frac{1}{2} \Lambda_{00} + \frac{1}{3} \varrho^1_{A,1} \varrho^1_{B,1}\Lambda_{11} + \frac{1}{3} \varrho^2_{A,2} \varrho^2_{B,2} \Lambda_{22} + \frac{1}{3}  \varrho^3_{A,3}  \varrho^3_{B,3} \Lambda_{33} \label{eq:Wern-dec} \\
& = & \begin{bmatrix}
\frac{1}{4} + \frac{  \varrho^3_{A,3}  \varrho^3_{B,3} }{6} & 0 & 0 & \frac{ \varrho^1_{A,1} \varrho^1_{B,1} - \varrho^2_{A,2} \varrho^2_{B,2}}{6} \\
 0 & \frac{1}{4} - \frac{  \varrho^3_{A,3}  \varrho^3_{B,3} }{6} & 
\frac{ \varrho^1_{A,1} \varrho^1_{B,1} + \varrho^2_{A,2} \varrho^2_{B,2}}{6}  & 0 \\
 0 & \frac{ \varrho^1_{A,1} \varrho^1_{B,1} + \varrho^2_{A,2} \varrho^2_{B,2} }{6}  & \frac{1}{4} - \frac{  \varrho^3_{A,3}  \varrho^3_{B,3} }{6} & 0 \\
 \frac{ \varrho^1_{A,1} \varrho^1_{B,1} - \varrho^2_{A,2} \varrho^2_{B,2} }{6} &  0 & 0 & \frac{1}{4} + \frac{  \varrho^3_{A,3}  \varrho^3_{B,3} }{6}
\end{bmatrix}
\nonumber
\eeqa
While the ``multiplicative'' part of $ \rho $ is nontrivial, the corresponding reduced densities are always completely mixed because of the cancellations occurring.
The eigenvalues of \eqref{eq:Wern-dec} are 
\begin{multline}
\left\{ \frac{1}{4} + \frac{1}{6} \left( - \varrho^1_{A,1} \varrho^1_{B,1} -\varrho^2_{A,2} \varrho^2_{B,2} -\varrho^3_{A,3} \varrho^3_{B,3} \right) , \; 
 \frac{1}{4} + \frac{1}{6} \left( - \varrho^1_{A,1} \varrho^1_{B,1} +\varrho^2_{A,2} \varrho^2_{B,2} +\varrho^3_{A,3} \varrho^3_{B,3} \right) , \; \right. \\
\left.  \frac{1}{4} + \frac{1}{6} \left(  \varrho^1_{A,1} \varrho^1_{B,1} -\varrho^2_{A,2} \varrho^2_{B,2} +\varrho^3_{A,3} \varrho^3_{B,3} \right) , \; 
 \frac{1}{4} + \frac{1}{6} \left(  \varrho^1_{A,1} \varrho^1_{B,1} +\varrho^2_{A,2} \varrho^2_{B,2} -\varrho^3_{A,3} \varrho^3_{B,3} \right) , \;  \right\}
\end{multline}
All four eigenvalues are $ \geqslant 0 $ for all admissible $ \vec{\varrho}_{A,p} , \,  \vec{\varrho}_{B,p} \in \mathbb{S}^2_{\leqslant \frac{1}{\sqrt{2}}} $.
In particular $ \rho $ of \eqref{eq:Wern-dec} is equal to $ \rho_{\rm Wer} $ provided one chooses $  \varrho^1_{A,1}= \varrho^2_{A,2} = \varrho^3_{A,3} $ and $  \varrho^1_{B,1} =-  \varrho^1_{A,1} $, $  \varrho^2_{B,2} =-  \varrho^2_{A,2} $ and $  \varrho^3_{B,3} =-  \varrho^3_{A,3} $. 
In this case, the eigenvalues never reach the critical points for a density ($0 $ and $+1 $).
If $ x= -\frac{2}{3} ( \varrho^1_{A,1} \varrho^1_{B,1}) = \frac{2}{3} ( \varrho^1_{A,1} )^2 $ then the allowed parameter span is $ x\in [ 0, \, \frac{1}{3} ]$, as is known for a separable Werner state.
In fact, $ x \geqslant 0 $ by definition (since $ x =\frac{2}{3} ( \varrho^1_{A,1} )^2 \geqslant 0 $), and $ x = \frac{2}{3} \left( \frac{1}{2} \right) $ when $ \varrho_{A,1}^1 $ reaches $ \pm \frac{1}{\sqrt{2} } $.
It is interesting to see what happens to the single terms of the convex sum.
For example
\beqan
\rho_{A,1} \otimes \rho_{B,1} & = & \left( \varrho^0 \lambda_0 + \varrho_{A,1}^1 \lambda_1 \right) \otimes  \left( \varrho^0 \lambda_0 + \varrho_{B,1}^1 \lambda_1 \right) \\
& = &  \varrho^0  \varrho^0 \Lambda_{00} +  \varrho^0 \varrho_{B,1}^1 \Lambda_{01} + \varrho_{A,1}^1\varrho^0 \Lambda_{10} + \varrho_{A,1}^1 \varrho_{B,1}^1\Lambda_{11} \\
& = & \frac{1}{2} \Lambda_{00} + \frac{\varrho_{B,1}^1}{\sqrt{2}}  \Lambda_{01} + \frac{\varrho_{A,1}^1}{\sqrt{2}} \Lambda_{10} + \varrho_{A,1}^1 \varrho_{B,1}^1\Lambda_{11}
\eeqan
is a well-defined density within the same boundary as $ \rho_{A,1} $, $ \rho_{B,1} $.
Instead if one considers the sum of two terms which cancels the one-qubit terms (for example ``1'' and ``4'' with equal weights $ w^1 = w^4 $)  
\beqan 
w^1 \rho_{A,1} \otimes \rho_{B,1} + w^4 \rho_{A,4} \otimes \rho_{B,4} & = & 2 w^1 \left(  \varrho^0  \varrho^0 \Lambda_{00} +  \varrho_{A,1}^1 \varrho_{B,1}^1\Lambda_{11} \right) \\
& = & 2 w^1 \left( \frac{1}{2} \Lambda_{00} - \frac{3}{2} x \Lambda_{11} \right) =  2 w^1 \left( \tilde{\rho} _{AB}  \right)
\eeqan
then $ \tilde{\rho} _{AB} $ is a well-defined density when $ \frac{1}{4} + \left( - \frac{3}{2} x \right) ^2 \leqslant 1 $ i.e., in the larger interval $ 0 \leqslant x \leqslant \frac{1}{\sqrt{3}} $.
The same bound is found on the other pairwise sums.
This is also the bound found by the trace-square inequalities \eqref{eq:tr-sq-red}.

Notice that choosing $  \varrho^1_{A,1}= \varrho^2_{A,2} = \varrho^3_{A,3} $ but instead $  \varrho^1_{B,1} =  \varrho^1_{A,1} $, $  \varrho^2_{B,2} =-  \varrho^2_{A,2} $ and $  \varrho^3_{B,3} =  \varrho^3_{A,3} $ and $ x= \frac{2}{3} ( \varrho^1_{A,1} \varrho^1_{B,1}) = - \frac{2}{3} ( \varrho^2_{A,2}  \varrho^2_{B,2}) $, in the same range of $ x$ one gets the different family:
\[
\rho_{\rm Wer}'= \begin{bmatrix} \frac{1+x}{4} & 0 & 0 & \frac{x}{2} \\
0 & \frac{1-x}{4} &0 & 0 \\
0 &0 &  \frac{1-x}{4} & 0 \\
\frac{x}{2} & 0 & 0 & \frac{1+x}{4} 
\end{bmatrix}
\]
which can also be obtained from $ \rho_{\rm Wer} $ by local unitary transformations and which leads to $  \rho_{\rm Bell} $ considered before for $ x=1$.

\subsection{A tripartite family}
\label{sec:ex-triparty1}
The one parameter density $ \rho = \varrho^{000} \Lambda_{000} + \varrho^{111} \Lambda_{111} =   \frac{1}{2 \sqrt{2} }  \Lambda_{000} + x \Lambda_{111} $, or written explicitly,
\beq
\rho= \frac{1}{2 \sqrt{2} } 
\begin{bmatrix} \frac{1}{2\,{\sqrt{2}}} & 0 & 0 & 0 & 0 & 0 & 0 & x \cr 
0 & \frac{1} {2\,{\sqrt{2}}} & 0 & 0 & 0 & 0 & x & 0 \cr 
0 & 0 & \frac{1} {2\,{\sqrt{2}}} & 0 & 0 & x & 0 & 0 \cr 
0 & 0 & 0 & \frac{1} {2\,{\sqrt{2}}} & x & 0 & 0 & 0 \cr 
0 & 0 & 0 & x & \frac{1} {2\,{\sqrt{2}}} & 0 & 0 & 0 \cr 
0 & 0 & x & 0 & 0 & \frac{1} {2\,{\sqrt{2}}} & 0 & 0 \cr 
0 & x & 0 & 0 & 0 & 0 & \frac{1} {2\,{\sqrt{2}}} & 0 \cr 
x & 0 & 0 & 0 & 0 & 0 & 0 & \frac{1}{2\,{\sqrt{2}}} \end{bmatrix}
\label{eq:rho-trip-fam1}
\eeq
has $ \tr{\rho^2} = \frac{1}{8} + x^2 $ and eigenvalues $ \left\{ \frac{1}{8}  \pm \frac{1}{2 \sqrt{2}} x \right\} $ each with multiplicity 4.
$ \rho $ is a density for $ -\frac{1}{2 \sqrt{2}} \leqslant x \leqslant \frac{1}{2 \sqrt{2}} $. 
From Corollary~\ref{cor:PPT-index-n}, this state is certainly PPT.  
Furthermore, the necessary conditions \eqref{eq:tr-norm-ineq-n} are satisfied as well as the realignment criterion \cite{Cheng1,Cheng2,Horodecki5}.
Still one may wonder if $\rho $ is separable and if an explicit mixture can be found and how.   
For $ x=0 $ (and near it) it is obviously separable. Once we have an explicit linear combination for $ \rho $, then this will provide an estimate of its separability interval, valid at least with respect to the mixture used.
All 1-qubit and 2-qubit reduced densities are in the completely random state.
The trick of pairwise canceling reduced densities does not generalize in a straightforward manner to multipartite qubits, but some variants of it can still be used.
For example the following 9 triples of operators 
\beqan
 \rho_{A,1} = \varrho^0 \lambda_0  & \qquad
 \rho_{B,1} = \varrho^0 \lambda_0 + \varrho^1_{B,1} \lambda_1 & \qquad
 \rho_{C,1} = \varrho^0 \lambda_0 + \varrho^1_{C,1} \lambda_1 \\
 \rho_{A,2} = \varrho^0 \lambda_0  & \qquad
 \rho_{B,2} = \varrho^0 \lambda_0 - \varrho^1_{B,1} \lambda_1 & \qquad
 \rho_{C,2} = \varrho^0 \lambda_0 - \varrho^1_{C,1} \lambda_1 
\eeqan
\beqan
 \rho_{A,3} = \varrho^0 \lambda_0 + \varrho^1_{A,3} \lambda_1 & \qquad
 \rho_{B,3} = \varrho^0 \lambda_0  & \qquad
 \rho_{C,3} = \varrho^0 \lambda_0 + \varrho^1_{C,3} \lambda_1 \\
 \rho_{A,4} = \varrho^0 \lambda_0 - \varrho^1_{A,3} \lambda_1 & \qquad
 \rho_{B,4} = \varrho^0 \lambda_0  & \qquad
 \rho_{C,4} = \varrho^0 \lambda_0 - \varrho^1_{C,3} \lambda_1 
\eeqan
\beqan
 \rho_{A,5} = \varrho^0 \lambda_0 + \varrho^1_{A,5} \lambda_1 & \qquad
 \rho_{B,5} = \varrho^0 \lambda_0 + \varrho^1_{B,5} \lambda_1 & \qquad
 \rho_{C,5} = \varrho^0 \lambda_0  \\
 \rho_{A,6} = \varrho^0 \lambda_0 - \varrho^1_{A,5} \lambda_1  & \qquad
 \rho_{B,6} = \varrho^0 \lambda_0 - \varrho^1_{B,5} \lambda_1 & \qquad
 \rho_{C,6} = \varrho^0 \lambda_0 
\eeqan
\beqan
 \rho_{A,7} = \varrho^0 \lambda_0 + \varrho^1_{A,7} \lambda_1 & \qquad
 \rho_{B,7} = \varrho^0 \lambda_0 + \varrho^1_{B,7} \lambda_1 & \qquad
 \rho_{C,7} = \varrho^0 \lambda_0 + \varrho^1_{C,7} \lambda_1 \\
 \rho_{A,8} = \varrho^0 \lambda_0 - \varrho^1_{A,7} \lambda_1 & \qquad
 \rho_{B,8} = \varrho^0 \lambda_0 - \varrho^1_{B,7} \lambda_1 & \qquad
 \rho_{C,8} = \varrho^0 \lambda_0 \\
 \rho_{A,9} = \varrho^0 \lambda_0 & \qquad
 \rho_{B,9} = \varrho^0 \lambda_0 & \qquad
 \rho_{C,9} = \varrho^0 \lambda_0 - \varrho^1_{C,7} \lambda_1 
\eeqan
with the choices of weights $ w^1 = w^2$, $ w^3=w^4$, $ w^5 = w^6 $ and $ w^7 = w^8 = w^9 $ such that $ w^1 + \ldots + w^9 = 1 $ gives
\[
\begin{split}
\rho & =  \frac{1}{2 \sqrt{2} }  \Lambda_{000} 
+ \frac{1}{\sqrt{2}} \left( 2 w^1  \varrho^1_{B,1}\varrho^1_{C,1} + w^7 \varrho^1_{B,7}\varrho^1_{C,7}  \right) \Lambda_{011} 
+ \frac{1}{\sqrt{2}} \left(2 w^3  \varrho^1_{A,3}\varrho^1_{C,3} + w^7 \varrho^1_{A,7}\varrho^1_{C,7} \right) \Lambda_{101} \\
& +  \sqrt{2} \left( w^5  \varrho^1_{A,5}\varrho^1_{B,5} + w^7 \varrho^1_{A,7}\varrho^1_{B,7} \right) \Lambda_{110} 
+w^7 \varrho^1_{A,7}\varrho^1_{B,7}\varrho^1_{C,7} \Lambda_{111} 
\end{split}
\]
In order to cancel the terms along $ \Lambda_{011}, \, \Lambda_{101}, \,   \Lambda_{110}$, we fix the parameters of the reduced operators as follows
\[
\varrho^1_{B,1} = \sqrt{\frac{(x)^{\frac{2}{3}} ( w^7 )^{\frac{1}{3}} }{2 w^1 }}, \qquad \varrho^1_{C,1} =-\varrho^1_{B,1} , \qquad 
\varrho^1_{A,3} = \sqrt{\frac{(x)^{\frac{2}{3}} ( w^7 )^{\frac{1}{3}} }{2 w^3 }}, \qquad \varrho^1_{C,3} =-\varrho^1_{B,3} ,
\]
\[
\varrho^1_{A,5} = \sqrt{\frac{(x)^{\frac{2}{3}} ( w^7 )^{\frac{1}{3}} }{ w^5 }}, \qquad \varrho^1_{B,5} =-\varrho^1_{A,5} , \qquad 
\varrho^1_{A,7} =\varrho^1_{B,7} =\varrho^1_{C,7} = \left( \frac{x}{w^7} \right)^{\frac{1}{3} }
\]
hence reobtaining \eqref{eq:rho-trip-fam1} for all $ x$. What is left to do is to control for which $ x $ there exist admissible weights $ w^r $ such that all the reduced operators are well-defined densities, i.e., $ | \varrho^1_{K, r} | \leqslant \frac{1}{\sqrt{2}} $, $ K = A,  B, C $, $ r = 1,3,5,7$.
Squaring the relations above, we obtain that $ x$ has to fulfill the 4 constraints 
\beqa
& -\frac{w^7}{2 \sqrt{2} } \leqslant x  \leqslant \frac{w^7}{2 \sqrt{2} } & \label{eq:constr-tripA1} \\
 \left( (x)^2 w^7 \right)^{\frac{1}{3} }  \leqslant w^1 &, \qquad 
\left( (x)^2 w^7 \right)^{\frac{1}{3} }  \leqslant w^3,  \qquad  & \left( (x)^2 w^7 \right)^{\frac{1}{3} }  \leqslant \frac{w^5}{2}  \label{eq:constr-tripA2}
\eeqa
One can rewrite the last three inequalities as
\beq
0 \leqslant x  \leqslant \left\{ \frac{(w^1)^3 }{w^7 } ; \,  \frac{(w^3)^3 }{w^7 }; \,  \frac{(w^5)^3 }  { 8 w^7 } \right\}
\label{eq:constr-tripA2-bis}
\eeq
Using the rule of thumb of selecting weights so that all three reduced densities constraints \eqref{eq:constr-tripA2-bis} become active simultaneously as we vary $x$, $ w^1=w^3=\frac{w^5}{2}$, one gets $ w^7 = \frac{1-8w^1}{3} $. As we vary $ w^1 $ in $ 0 \leqslant w^1  \leqslant \frac{1}{8} - \epsilon $ , $ \epsilon > 0$, the inequalities \eqref{eq:constr-tripA1} and \eqref{eq:constr-tripA2-bis} give a trade off. A quick numerical search shows that $ -0.050 \leqslant x  \leqslant 0.050 $ is the largest common range satisfying the constraints one can achieve, corresponding to $ w^1 = 0.0715$.
In fact, in correspondence of these values we have that all 1- and 2- qubit densities have trace squares between $ 0.99 $ and $1$ (while $ \tr{\rho^2} =0.125 $).

One may wonder if the bound found depends on the linear combination chosen or less. In order to verify this, it is possible to use the alternative family of densities:
\beqan
 \rho_{A,1} = \varrho^0 \lambda_0  & \qquad
 \rho_{B,1} = \varrho^0 \lambda_0  & \qquad
 \rho_{C,1} = \varrho^0 \lambda_0 - \varrho^1_{C,1} \lambda_1 \\
 \rho_{A,2} = \varrho^0 \lambda_0 + \varrho^1_{A,2} \lambda_1 & \qquad
 \rho_{B,2} = \varrho^0 \lambda_0 + \varrho^1_{B,2} \lambda_1 & \qquad
 \rho_{C,2} = \varrho^0 \lambda_0  \\
 \rho_{A,3} = \varrho^0 \lambda_0 - \varrho^1_{A,2} \lambda_1 & \qquad
 \rho_{B,3} = \varrho^0 \lambda_0 - \varrho^1_{B,2} \lambda_1 & \qquad
 \rho_{C,3} = \varrho^0 \lambda_0  \\
 \rho_{A,4} = \varrho^0 \lambda_0 + \varrho^1_{A,4} \lambda_1 & \qquad
 \rho_{B,4} = \varrho^0 \lambda_0 + \varrho^1_{B,4} \lambda_1  & \qquad
 \rho_{C,4} = \varrho^0 \lambda_0 + \varrho^1_{C,4} \lambda_1 \\
 \rho_{A,5} = \varrho^0 \lambda_0 + \varrho^1_{A,4} \lambda_1 & \qquad
 \rho_{B,5} = \varrho^0 \lambda_0 + \varrho^1_{B,4} \lambda_1 & \qquad
 \rho_{C,5} = \varrho^0 \lambda_0 - \varrho^1_{C,4} \lambda_1
\eeqan
corresponding to 
\beqan
\rho  & = & \frac{1}{2 \sqrt{2} }  \Lambda_{000} 
+ \left( \frac{- w^1  \varrho^1_{C,1} }{2} + w^4  \varrho^1_{C,4} \right) \Lambda_{001} 
+  \sqrt{2} \left( w^2  \varrho^1_{A,2}\varrho^1_{B,2} + w^4 \varrho^1_{A,4}\varrho^1_{B,4} \right) \Lambda_{110}+  \\
&& + 2 w^4 \varrho^1_{A,4}\varrho^1_{B,4}\varrho^1_{C,4} \Lambda_{111} 
\eeqan
Choosing $ w^3 = w^2 $, $ w^5=w^4 $ and 
\beqan
 \varrho^1_{C,1} & = & -\frac{2}{w^1} \left( \frac{x ( w^4 )^2}{2} \right)^{\frac{1}{3}} , \\
 \varrho^1_{A,2} & = & - \varrho^1_{B,2} = \sqrt{ \frac{1}{w^2} \left( \frac{ (x)^2 w^4 } {4} \right)^{\frac{1}{3} }}, \\
 \varrho^1_{A,4}& = & \varrho^1_{B,4}=\varrho^1_{C,4}= \left( \frac{ (x)^2} {4  (w^4)^2} \right)^{\frac{1}{3} }
\eeqan
this convex combination has components that are all densities when the three inequalities are satisfied:
\[
- \frac{(w^1)^3 }{8 \sqrt{2} ( w^4)^2 }  \leqslant x  \leqslant \frac{(w^1)^3 }{8 \sqrt{2} ( w^4)^2 },\qquad 
- \sqrt{ \frac{(w^2)^3 }{2 w^4 } }  \leqslant x  \leqslant \sqrt{ \frac{(w^2)^3 }{2 w^4 } } , \qquad 
- \frac{w^4}{\sqrt{2} }   \leqslant x  \leqslant \frac{w^4}{\sqrt{2} } .
\]
A parametric search gives that $ x$ is separable in $ - 0.1166 \leqslant x \leqslant 0.1166 $ in correspondence for example of $ w^1 =0.33$, $ w^2=0.17 $ and $ w^4 =0.165$.
Hence just like the mixture representing a density is nonunique, also the separability range changes with the convex combination chosen.
No conclusion is drawn about separability in $ 0.1166 \leqslant | x | \leqslant \frac{1}{2 \sqrt{2}} $, although this density is probably separable on the whole range.

\subsection{Another tripartite biseparable family}
\label{sec:ex-triparty2}
Consider the 3-qubit state $ \rho = \varrho^{jkl}  \Lambda_{jkl} $ where $ \varrho^{000} = \frac{1}{2\sqrt{2} }$ and 
\beq
\begin{split}
& \varrho^{031}=\varrho^{033}=\varrho^{103}=\varrho^{111}=\varrho^{133}=\varrho^{303}=\varrho^{310}=\varrho^{313}=\varrho^{330}=\varrho^{331}=x  \\
& \varrho^{011}=\varrho^{013}=\varrho^{101}=\varrho^{110}=\varrho^{130}=\varrho^{301}=-x  \\
& \varrho^{jkl} =0 \qquad \text{otherwise} .
\end{split}
\label{eq:bisep-varrho}
\eeq
Expanding explicitly
\[
\rho=\frac{1}{2 \sqrt{2} }
\begin{bmatrix} \frac{{\sqrt{2}} + 12\,x}{4} & x & x & -x & x & -x & -x & x \cr
 x & \frac{{ \sqrt{2}} - 4\,x}{4} & -x & x & -x & -3\,x & x & -x \cr
 x & -x & \frac{{\sqrt{2}} -      4\,x}{4} & -3\,x & -x & x & x & -x \cr
 -x & x & -3\,x & \frac{{\sqrt{2}} - 4\,x}
   {4} & x & -x & -x & x \cr 
x & -x & -x & x & \frac{{\sqrt{2}} - 4\,x}{4} & x & -3\,x & -x \cr
 -x & -3\,x & x & -x & x & \frac{{\sqrt{2}} - 4\,x}{4} & -x & x \cr 
    -x & x & x & -x & -3\,x & -x & \frac{{\sqrt{2}} - 4\,x}{4} & x \cr
 x & -x & -x & x &  -x & x & x & \frac{{\sqrt{2}} + 12\,x}{4} \cr 
 \end{bmatrix} .
\]
The trace norm for this density is $ \tr{\rho^2} = \frac{1}{8} + 16 x^2 $ and the eigenvalues $ \left\{ \frac{1}{8} \pm \sqrt{2} x \right\} $ each with multiplicity 4. Hence $ \rho $ is a well defined density matrix for $ -\frac{1}{8 \sqrt{2} } \leqslant x \leqslant \frac{1}{8 \sqrt{2} } $ and it is mixed in the entire interval (completely mixed for $ x=0 $).
When $ x= \frac{1}{8 \sqrt{2}} $ we recover the example of Eq. (6) of \cite{Bennett1} $ \rho_{\rm UPB} = \frac{1}{4} \left( \openone_8 - \sum_{j=1}^4 \ket{\psi_j} \bra{\psi_j } \right) $, with $ \ket{\psi_j} $ an Unextendible Product Basis (UPB) for 3-qubit states: $ \ket{\psi_j} = \ket{01+}, \ket{1+0},\ket{+01}, \ket{---} $ (where $\ket{\pm} = \frac{1}{\sqrt{2}} \left( \ket{0} \pm \ket{1} \right) $). $ \rho_{\rm UPB} $ is known to be entangled and PPT.
Again, this last fact is simply verified by noticing that no index ``2'' appear in \eqref{eq:bisep-varrho}, hence Corollary~\ref{cor:PPT-index-n} holds for all $ x$.

Notice that for this example the necessary conditions \eqref{eq:tr-norm-ineq-n} are fulfilled for all $x$, but that $ \rho$ is separable only in some proper subinterval $ [ - x_c , \, x_c ] $ around the complete mixing state.

Also in this case it is possible to obtain a linear convex combination of tensor products of trace 1 Hermitian matrices corresponding to $ \rho$. This allows to estimate the separability interval with respect to the particular mixture chosen by imposing that each matrix of each tensor product be a well-defined density.
One notices first that 1-qubit reduced densities are all completely mixed ($ \varrho^{00l} = \varrho^{0k0} = \varrho^{j00} =0 $). 
Hence for each nonzero 2-qubit term a cancellation on the corresponding reduced terms must occur, as in the Werner state parameterization.
On the contrary, 2-qubit reduced densities are not completely random and it is straightforward to observe that all of the nonzero 2-qubit terms correspond to ``reductions'' of the 4 nonzero 3-qubit terms: for example $ \varrho^{133} \neq 0 $ implies $ \varrho^{033} \neq 0 $ $ \varrho^{103} \neq 0 $ $ \varrho^{130} \neq 0 $. 
If we use the same scheme of Section \ref{sec:ex-triparty1} to produce the 3 party terms, then we obtain a convex combination of 36 triples of two level densities which, with the naive choice of all equal weights, guarantees separability only in a tiny interval: $ -0.006 \leqslant x \leqslant 0.006$ (the realignment criterion detects entanglement only close to $ \frac{1}{8 \sqrt{2} } $). 
The explicit expression of the components is available upon request.
Notice that the linear combination obtained connects $ \rho_{\rm UPB} $ with $ \varrho^{000} \Lambda_{000} $.
From the classification of \cite{Acin1}, it is known that the biseparable entangled regions have a border in common with the set of separable states.
Since this last set is convex, the linear combination has a unique crossing point $ x_{c'} $ (plus, specularly, $ - x_{c'} $).

Managing densities and weights becomes rapidly cumbersome, but one can expect the problem to admit an algorithmic formulation via (tractable) convex optimization schemes.

\subsection{Building NPT entangled states}
\label{sec:ex-triparty3}
\begin{figure}[t!]
\begin{center}
\subfigure[] {
 \includegraphics[width=4.5cm]{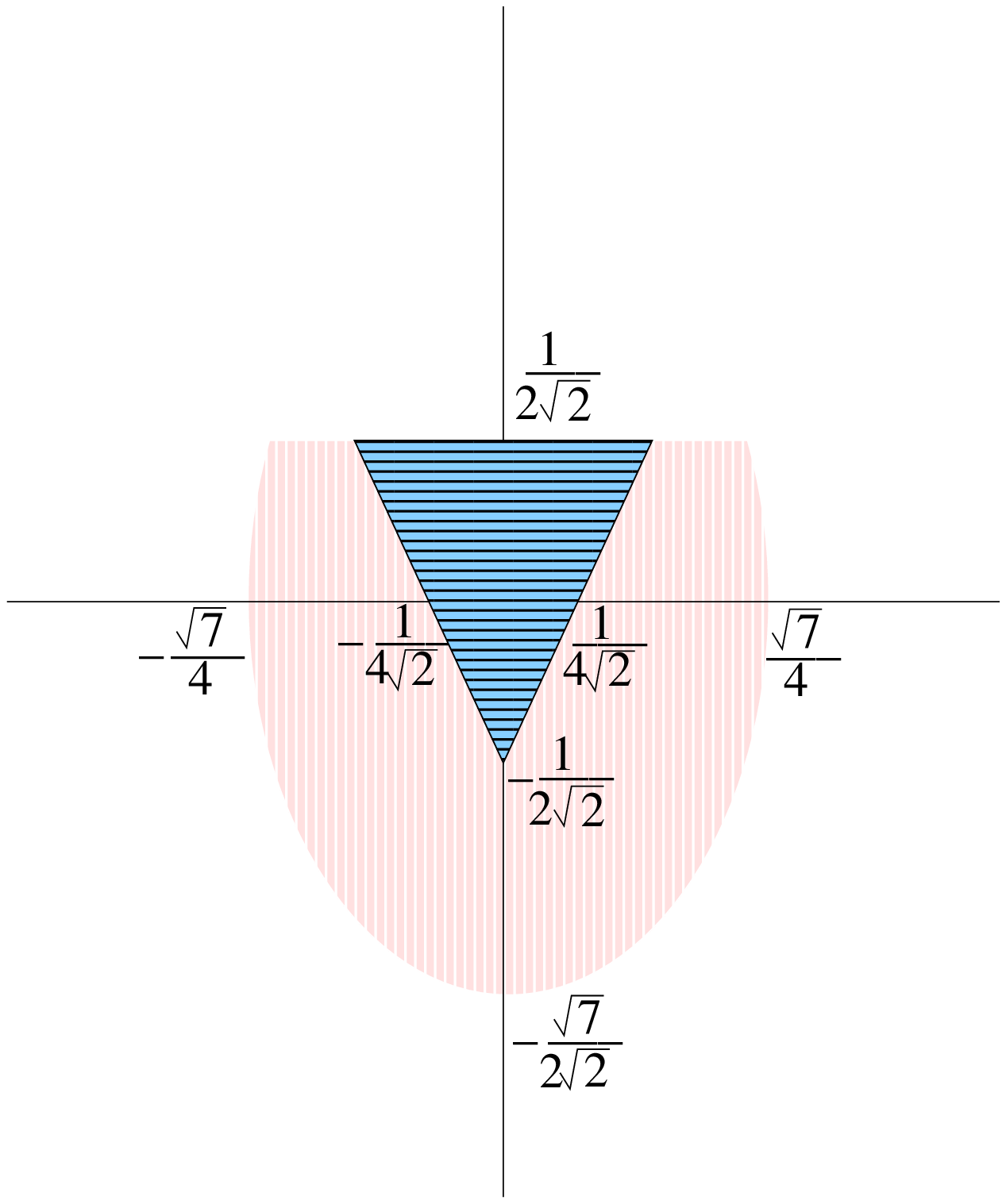} }
\subfigure[] {
 \includegraphics[width=4.5cm]{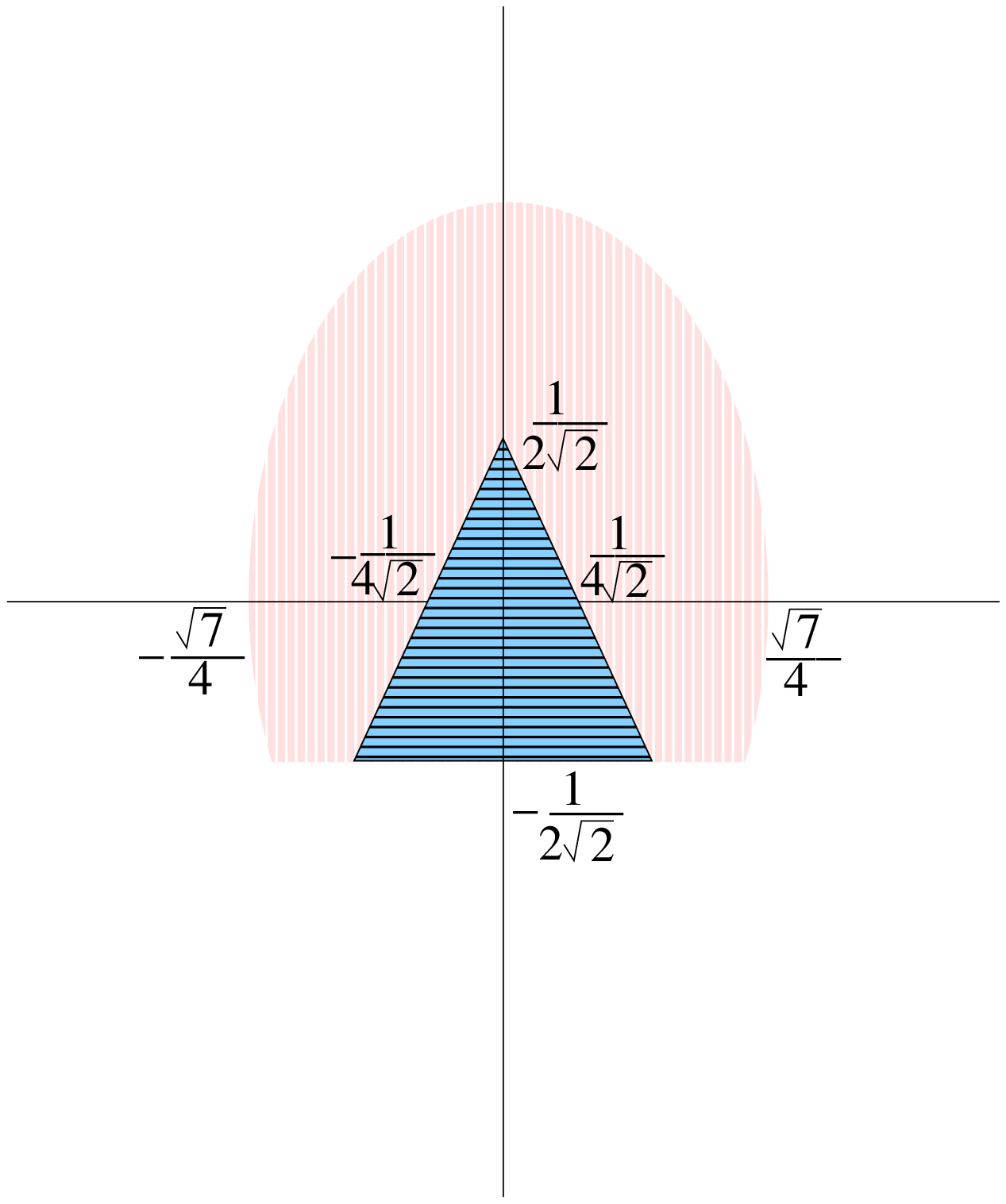} }
\subfigure[] {
 \includegraphics[width=4.5cm]{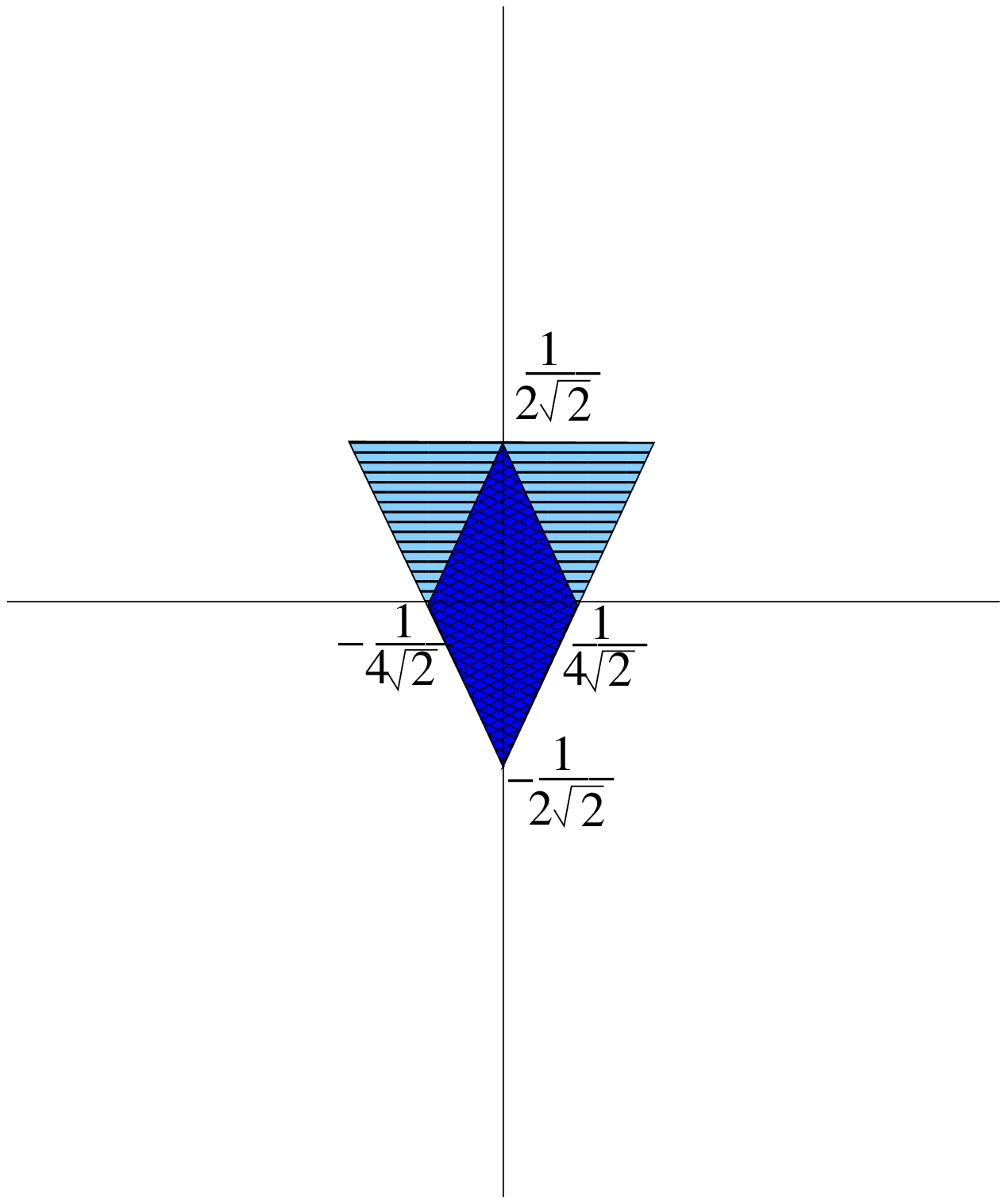} }
\caption{$ (x,y) $ parameter space for the example of Section~\ref{sec:ex-triparty3}. In (a) and (b) the triangle represents the admissible $ (x,y)$ pairs for $ \rho_{AB-C} $ and $ \rho_{AB-C}^{T_1} $ respectively. Hence $ (x,y)$ belonging to the complement of the rhomb in the triangle of (c) gives NPT entangled densities.}  
\label{fig:sep-ex1}
\end{center}
\end{figure}

NPT entanglement (i.e., entanglement detectable by means of the PPT test) is obviously the simplest to manipulate for the purposes of building entanglement between subsystems.
There are by now standard methods to do that, typically starting from particular ket states or taking linear combinations of known singlets with product states.
The aim of this example is just to show that such a construction is also easy in the formalism we are proposing, and the states one obtains are not the standard one normally considered in the literature (GHZ, W states, etc.) \footnote{Although they may very well be reconducible to the standard states of a classification as that of \cite{Acin1} by means of local unitaries and classical communication.}.
For example, assume that one wants to have three qubits with A and B NPT entangled, C satisfying the PPT property.
One possible solution is the following state
\beq
\rho_{AB-C} = \frac{1}{2 \sqrt{2}}\Lambda_{000} + x \Lambda_{122} + x \Lambda_{212} + y \Lambda_{330}
\label{eq:rhoAB-C}
\eeq
For $ \rho_{AB-C} $ we have $ \tr{\rho_{AB-C}^2} = \frac{1}{8} + 2 x^2 + y^2 $ and eigenvalues $ \frac{1}{8} \left( 1 - 2 \sqrt{2} y \right) $ of multiplicity 4 and $ \frac{1}{8} \left( 1 + 2 \sqrt{2}\left(  y \pm 2 x \right)  \right) $ each of multiplicity 2.
$ \rho^{T_1} $ and $ \rho^{T_2} $ instead have eigenvalues $ \frac{1}{8} \left( 1 + 2 \sqrt{2} y \right) $ with multiplicity 4 and $ \frac{1}{8} \left( 1 - 2 \sqrt{2}\left(  y \pm 2 x \right) \right)  $ each of multiplicity 2.
$ \rho_{AB-C} $ is a state in the triangle of Figure~\ref{fig:sep-ex1} (a), while $ \rho_{AB-C}^{T_1} $ (or $ \rho_{AB-C}^{T_2} $) has positive eigenvalues in the flipped triangle of Figure~\ref{fig:sep-ex1} (b). 
Hence the area outside the rhomb of Figure~\ref{fig:sep-ex1} (c) gives the pairs $(x,y) $ leading to a NPT entangled state.
Notice that dropping any of the three terms in \eqref{eq:rhoAB-C} the density becomes PPT.

\section{Conclusion}

For multiparty qubit densities, the tensorial formalism proposed, although intuitively simple and certainly not so original, has several advantages which are now summarized.
\begin{itemize}
\item First and foremost: it provides a lot of inside in the correlation pattern between subsystems and between ``groups'' of subsystems, as it is known from the literature \cite{Fano2,Mahler1}.
Hence it may be useful in the construction of {\em quantum networks}.
\item It allows to explain the concept of reduced density in terms of the affine parameterization and the tensor itself is given by all ``degrees of reduction'' down to the single qubit densities in an unambiguous way.
\item It also allows to give a geometric picture of the density in terms of juxtaposition of affine Bloch spheres.
Thanks to the affine parameterization, also reduced densities enter into the picture. For example the degree of purity of the $n$-party densities is given in terms of the radius of the sphere of the corresponding dimension via the standard trace norm.
Such norm itself can be intended as a Tsallis entropy for a particular choice of the index or as a quadratic Casimir invariant for the set of density operators.
\item It allows to construct a density with the desired correlations between its subsystems. In particular also quantum correlations of the NPT type.
\item It provides an intuitive interpretation of both NPT and bound entanglement: they correspond to linear combinations of tensor products in which at least one of the factors in one of the products is not a well-defined density in the sense that its Bloch vector is too big in norm.
\end{itemize}

In conclusion, we hope that the formalism put forward in this paper may help not only in the understanding of quantum correlations in multipartite systems, but also in their concrete and systematic engineering.

\appendix

\section{Tensor of coherences for two qubit states}
\label{app:two-spin-comp}
If $ (\rho)_{pq} $ $ p, \, q = 1, \ldots 4 $, are the elements of the density operator $ \rho $ of Section~\ref{sec:two-qubits}, the tensor of coherences $ \varrho^{jk}= \tr{\rho \Lambda_{jk}} $, $ j, \, k = 0, \ldots, 3$, corresponding to it has components:
\beqan
\varrho^{00} & = &  \frac{1}{2} \left( (\rho)_{11} + (\rho)_{22} + (\rho)_{33} +  (\rho)_{44} \right)   \\
\varrho^{01} & = &   {\rm Re }[(\rho)_{12}] + {\rm Re }[(\rho)_{34}]  \\
\varrho^{02} & = & - {\rm Im }[(\rho)_{12}] - {\rm Im }[(\rho)_{34}] \\
\varrho^{03} & = &  \frac{1}{2} \left( (\rho)_{11} - (\rho)_{22} + (\rho)_{33} - (\rho)_{44} \right) \\
\varrho^{10} & = &   {\rm Re }[(\rho)_{13}] + {\rm Re }[(\rho)_{24}] \\
\varrho^{20} & = & - {\rm Im }[(\rho)_{13}] - {\rm Im }[(\rho)_{24}] \\
\varrho^{30} & = & \frac{1}{2} \left( (\rho)_{11} + (\rho)_{22} - (\rho)_{33} -  (\rho)_{44} \right)  \\
\varrho^{11} & = &   {\rm Re }[(\rho)_{14}] + {\rm Re }[(\rho)_{23}]  \\
\varrho^{12} & = & - {\rm Im }[(\rho)_{14}] + {\rm Im }[(\rho)_{23}]  \\
\varrho^{13} & = &   {\rm Re }[(\rho)_{13}] - {\rm Re }[(\rho)_{24}]  \\
\varrho^{21} & = & - {\rm Im }[(\rho)_{14}] - {\rm Im }[(\rho)_{23}] \\
\varrho^{22} & = & - {\rm Re }[(\rho)_{14}] + {\rm Re }[(\rho)_{23}]  \\
\varrho^{23} & = & - {\rm Im }[(\rho)_{13}] + {\rm Im }[(\rho)_{24}] \\
\varrho^{31} & = &   {\rm Re }[(\rho)_{12}] - {\rm Re }[(\rho)_{34}] \\
\varrho^{32} & = & - {\rm Im }[(\rho)_{12}] + {\rm Im }[(\rho)_{34}] \\
\varrho^{33} & = &  \frac{1}{2} \left( (\rho)_{11} - (\rho)_{22} - (\rho)_{33} +  (\rho)_{44} \right) 
\eeqan

\bibliographystyle{abbrv}

\end{document}